\documentclass[conference]{IEEEtran}
\IEEEoverridecommandlockouts
\usepackage{enumitem}
\usepackage{cite}
\usepackage{amsmath,amssymb,amsfonts}
\usepackage{algorithmic}
\usepackage{graphicx}
\usepackage{textcomp}
\usepackage{xcolor}
\usepackage{booktabs}
\usepackage{multirow}
\def\BibTeX{{\rm B\kern-.05em{\sc i\kern-.025em b}\kern-.08em
    T\kern-.1667em\lower.7ex\hbox{E}\kern-.125emX}}
\begin{document}

\title{A Standardized Pipeline for Colon Nuclei Identification and Counting Challenge\\
}

\author{Jijun Cheng$^\dagger$, Xipeng Pan$^\dagger$, Feihu Hou, Bingchao Zhao, Jiatai Lin, Zhenbing Liu, Zaiyi Liu, Chu Han
\thanks{Chu Han, Xipeng Pan, Jiatai Lin, Bingchao Zhao and Zaiyi Liu are with the Department of Radiology, Guangdong Provincial People’s Hospital, Guangdong Academy of Medical Sciences, Guangzhou, Guangdong, 510080, China.}
\thanks{Jijun Cheng, Xipeng Pan, Feihu Hou and Zhenbing Liu are with the School of Computer Science and Information Security, Guilin University of Electronic Technology, Guilin, Guangxi, China.}
\thanks{$^\dagger$ Equal contribution.}
\thanks{Corresponding author: Han Chu (email: hanchu@gdph.org.cn)}
}

\maketitle

\begin{abstract}
Nuclear segmentation and classification is an essential step for computational pathology. TIA lab from Warwick University organized a nuclear segmentation and classification challenge (CoNIC)~\cite{graham2021conic} for H\&E stained histopathology images in colorectal cancer with two highly correlated tasks, nuclei segmentation and classification task and cellular composition task. There are a few obstacles we have to address in this challenge, 1) limited training samples, 2) color variation, 3) imbalanced annotations, 4) similar morphological appearance among classes. To deal with these challenges, we proposed a standardized pipeline for nuclear segmentation and classification by integrating several pluggable components. First, we built a GAN-based model to automatically generate pseudo images for data augmentation. Then we trained a self-supervised stain normalization model to solve the color variation problem. Next we constructed a baseline model HoVer-Net with cost-sensitive loss to encourage the model pay more attention on the minority classes. According to the results of the leaderboard, our proposed pipeline achieves 0.40665 mPQ+ (Rank 49th) and 0.62199 r2 (Rank 10th) in the preliminary test phase. 
\end{abstract}

\begin{IEEEkeywords}
Nuclear segmentation and classification, Computational pathology, Imbalanced annotations, Data augmentation, Stain normalization
\end{IEEEkeywords}

\section{Introduction}
Automated nuclei segmentation and classification in whole slide images can help objective assessment of tumor microenvironment~\cite{bera2019artificial}. Many previous works have been proposed for this task in different technical perspectives, multi-task learning~\cite{zhao2020triple,graham2019hover}, weakly-supervised learning~\cite{qu2020weakly} and etc~\cite{zhang2022ddtnet}. Recently, TIA lab from Warwick University organized a nuclei segmentation and classification challenge (CoNIC)~\cite{graham2021conic} on Grand Challenge Platform for colorectal cancer. This challenge is based on the recently proposed Lizard dataset~\cite{graham2021lizard}, which is the largest nuclei segmentation and classification dataset for colorectal cancer with 495,179 nuclear instances with classification labels. Different from the other existing datasets like Pannuke~\cite{gamper2019pannuke}, Lizard only focuses on colorectal cancer and it further divides inflammatory nuclei into several sub-categories, lymphocyte, plasma, neutrophil or eosinophil. There are two tasks in this challenge, the segmentation and classification task as well as the cellular composition task.

\begin{figure}[t]
	\centering
	\includegraphics[width=.99\linewidth]{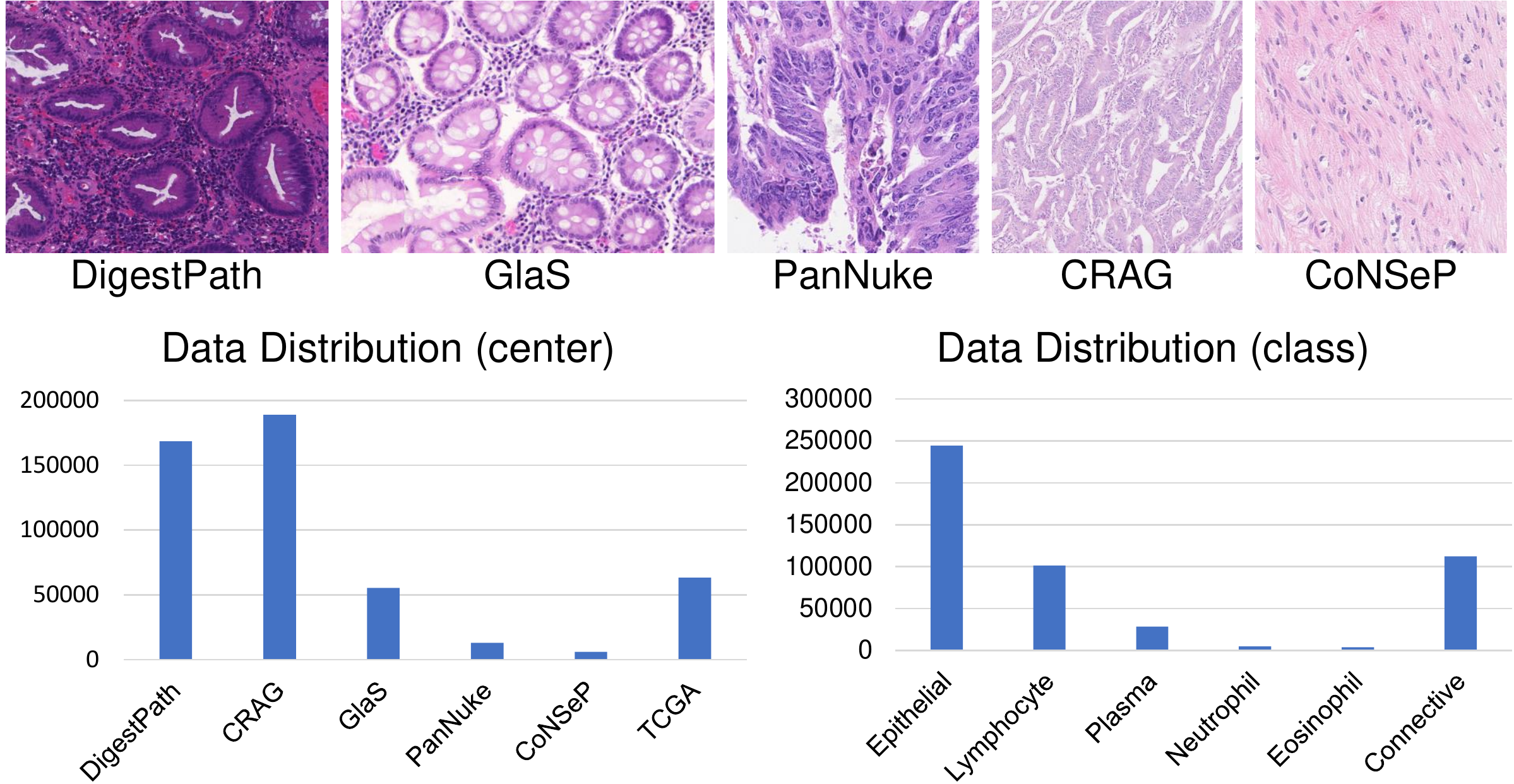}
	\caption{Data distribution of Lizard dataset.}
	\label{fig:dataset}
\end{figure}
\begin{figure*}[ht]
	\centering
	\includegraphics[width=.9\linewidth]{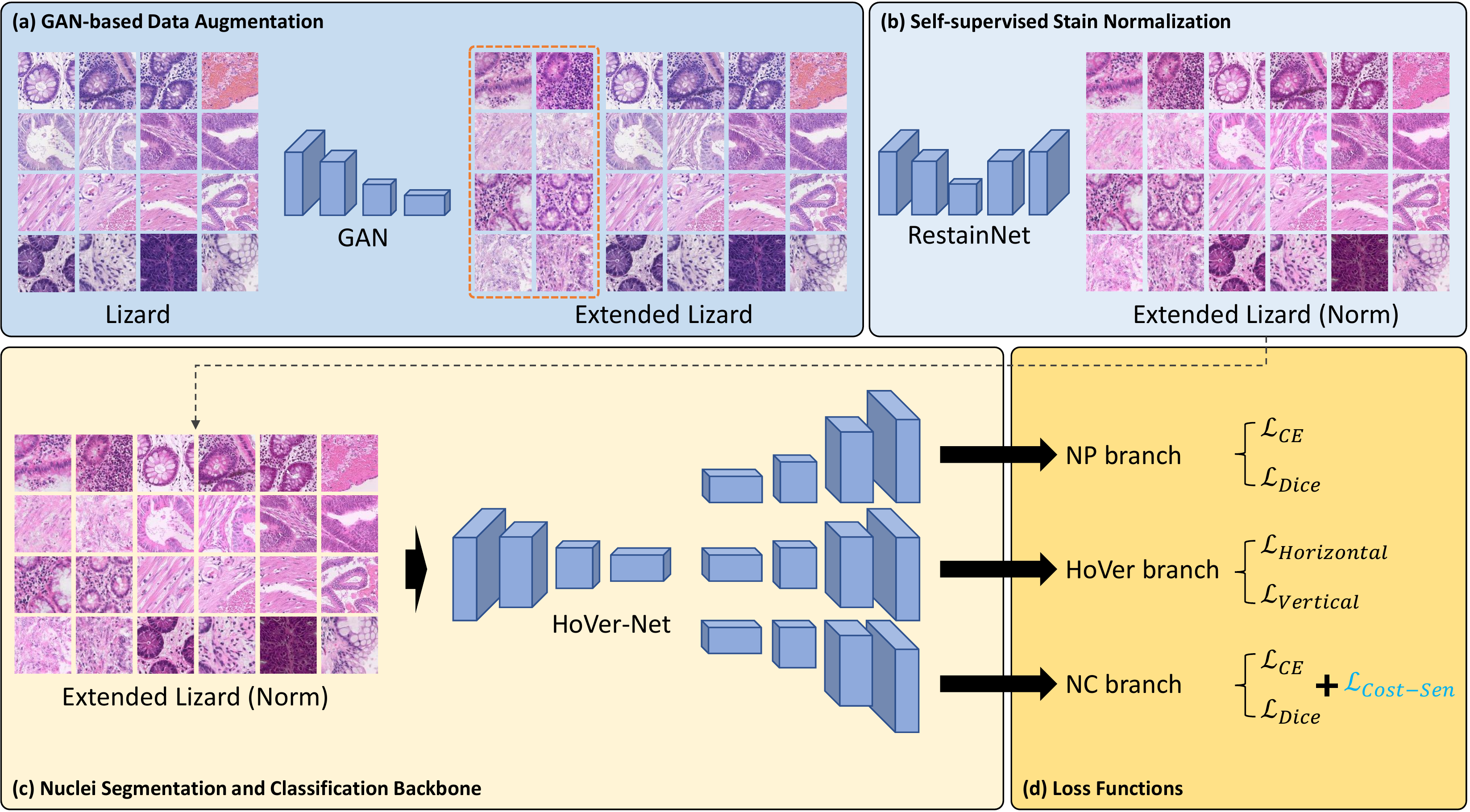}
	\caption{Overview of our proposed standardized pipeline for nuclei segmentation and classification.}
	\label{fig:overview}
\end{figure*}
Fig.~\ref{fig:dataset} shows the data distribution and some examples of Lizard dataset. We face some inherent difficulties. 1) The classification labels are extremely imbalanced. This challenge comprised of 244,563 epithelials, 101,413 lymphocytes, 28,466 plasmas, 4,824 neutrophils and 3,604 eosinophils and 112,309 connectives. The number of epithelials is almost 70 times larger than the number of eosinophils. Lack of training samples make neural network models hard to train a well feature representation for the minority classes. 2) Since the images are from multiple datasets in multiple centers, there exist some color variation because of different staining protocols or scanners. 3) Even it is the largest dataset so far for only one specific cancer, we can say the samples are still not sufficient for training a robust model because of the heterogeneous of malignant tumor. 4) The morphological appearances among different types are similar, such as lymphocytes and plasma cells, tumor epithelial nuclei and normal epithelial nuclei. Such difficulties not only exist in CoNIC challenge, they are also the most common challenges in the nuclei segmentation and classification task for all the cancers. How to solve them might be the key to build a stable and general model.

\begin{figure}[t]
	\centering
	\includegraphics[width=.99\linewidth]{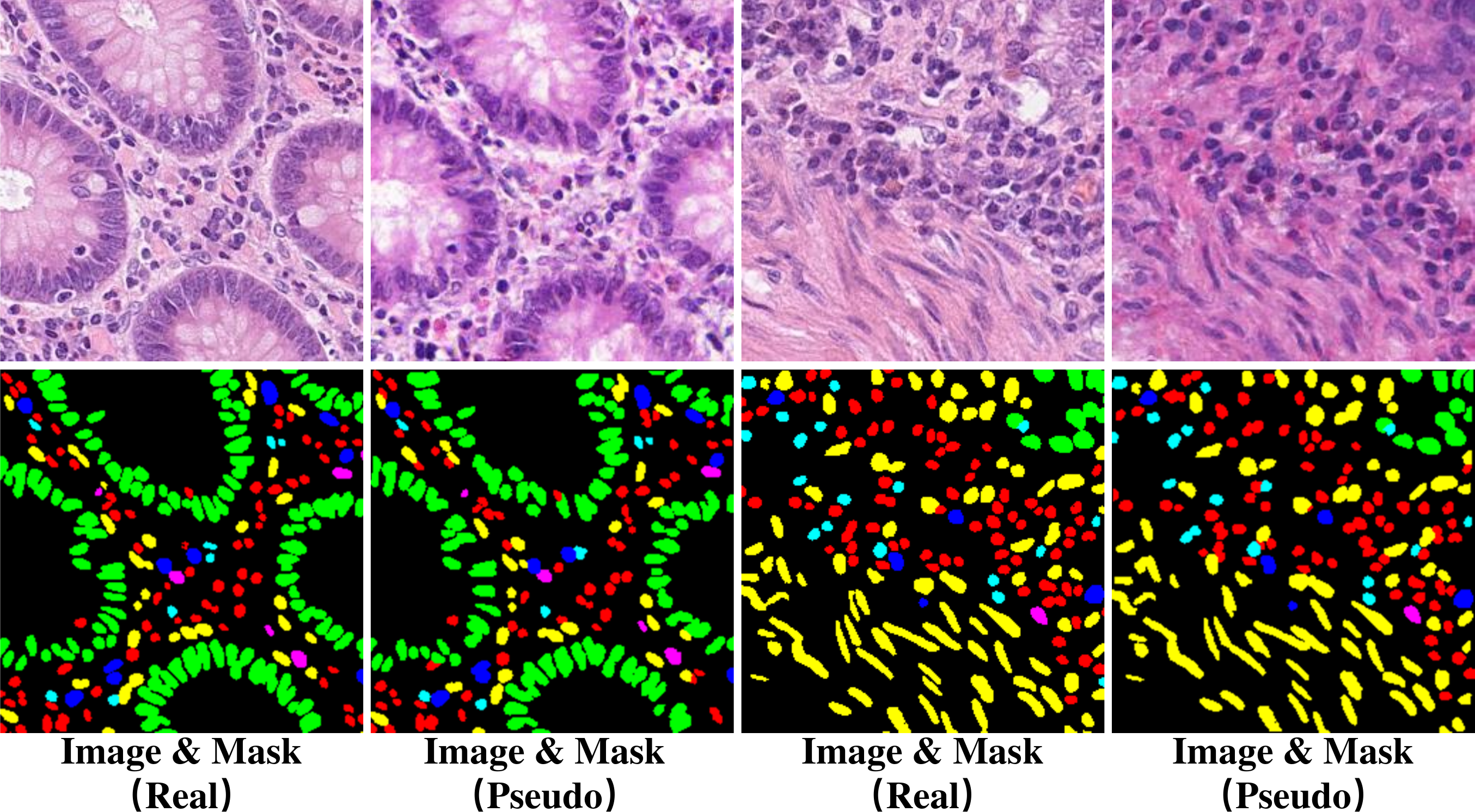}
	\caption{Generated pseudo images and pseudo labels.}
	\label{fig:case}
\end{figure}
In this paper, we achieve nuclei segmentation and classification by a systematic design other than developing a single model. With the help of our previous works in digital pathology, we proposed a systematic and standardized pipeline for nuclei segmentation and classification by integrating several independent components, including data augmentation, stain normalization, CNN backbone and loss functions. First, we applied a GAN-based model~\cite{cheng2021deep} to generate paired pseudo masks and images to extend the training set. Next, we introduced a self-supervised stain normalization model~\cite{zhao2022restainnet} to make the color style consistent. Finally, we applied a strong baseline model HoVer-Net~\cite{graham2019hover} as the backbone with an additional cost-sensitive loss~\cite{zhou2005training} to tackle the class imbalanced problem. Our standardized pipeline is pluggable and all the components can be replaced by more superior ones. On the leaderboard in the preliminary test phase, we can see all the components have proved its effectiveness according to the quantitative results. Finally, our proposed standardized pipeline achieves 0.40665 mPQ+ (Rank 49th) and 0.62199 r2 (Rank 10th) in the preliminary test phase.

\begin{table*}[t]
	\centering
	\caption{Quantitative Results - Preliminary Test Phase}
	\label{tab:task1}
	\begin{tabular}{c|ccc|ccccccc}
		\toprule[2pt]
		\multicolumn{11}{c}{\textbf{TASK 1 - Preliminary Test Phase (PQ)}}  \\ \hline
		\textbf{Model}  & \textbf{Data} & \textbf{Stain} & \textbf{Loss}& \textbf{mPQ+}  & \textbf{pla} & \textbf{neu} & \textbf{epi} & \textbf{lym} & \textbf{eos} & \textbf{con} \\ \hline
		\multirow{3}{*}{HoVer-Net} &- &- &- &0.2874 &0.2703 &0.0321 &0.4298 &0.4195 &0.2789 &0.2938 \\
		&- &$\checkmark$ &- &0.3889 &0.4353 &0.1855  &0.5234 &0.4115 &0.3794 &\textbf{0.3985}  \\
		&$\checkmark$ &$\checkmark$ &$\checkmark$ & \textbf{0.4067} & \textbf{0.4403} & \textbf{0.2005} & \textbf{0.5368} & \textbf{0.4394} & \textbf{0.4301} & 0.3928 \\ 
		\hline
		
		\multicolumn{11}{c}{\textbf{TASK 2 - Preliminary Test Phase (r2)}} \\ \hline
		\textbf{Model} & \textbf{Data} & \textbf{Stain} & \textbf{Loss} & \textbf{overall} & \textbf{pla} & \textbf{neu} & \textbf{epi} & \textbf{lym} & \textbf{eos} & \textbf{con} \\
		\hline
		\multirow{4}{*}{HoVer-Net}   &- &- &- & 0.0756 & 0.4742 & 0.2843 & 0.5531 &-1.8573 & 0.6670 &0.3346 \\
		&- &$\checkmark$ &- &0.4369 &0.7570 &0.5104 &0.6862 &-0.4704 &0.7554 &\textbf{0.3831}  \\
		&- &$\checkmark$ &$\checkmark$ &0.4514 &0.7485 &0.5048 &0.6912 &-0.3870 &\textbf{0.7939} &0.3570 \\
		&$\checkmark$ &$\checkmark$ &- &\textbf{0.6220} &\textbf{0.8022} &\textbf{0.8265} &\textbf{0.7858} &\textbf{0.3562} &0.7191 &0.2420 \\
		\toprule[2pt]
	\end{tabular}
\end{table*}

\section{Methodology}
Fig.~\ref{fig:overview} demonstrates the standardized pipeline for nuclei segmentation and classification. It includes four components, GAN-based data augmentation, self-supervised stain normalization, CNN backbone and loss functions.

\subsection{GAN-based Data Augmentation}
Besides the most common data augmentation ways like rotation, flipping and etc, we introduced our previously proposed GAN-based model for data augmentation~\cite{cheng2021deep}. This model used the masks from Lizard dataset as the input image and generated pseudo histopathology images. With a well trained generator, we randomly construct new pseudo labels and generated pseudo pathological images.

In order to avoid inputting exactly the same masks to the generator while keeping the real distribution of the nuclei, we introduced some slight vibrations for each real mask. First of all, we extract every nuclear instance in the semantic segmentation masks in the Lizard dataset. For each nuclear instance, we introduced random $x$ and $y$ axis translations within the range of (-3,3) pixels and a random rotation within the angle of $-3^{\circ}$ and $+3^{\circ}$ while keeping the class and the shape unchanged. Such slight changes aim to conforms to the real distribution of cells in the real histopathological images. We generated 1000 pseudo images and masks pairs for data augmentation. Fig.~\ref{fig:case} shows the generated pseudo pathological images using pseudo labels.

\subsection{Self-supervised Stain normalization}
Color variation problem may greatly hurt the model generalization ability. Since this challenge does not allow to use other public pathology datasets. We introduced our previously proposed self-supervised stain normalization approach RestainNet~\cite{zhao2022restainnet}. We used the CoNSeP dataset, which is a part of Lizard dataset, to train our stain normalization model. Since RestainNet is a self-supervised model, paired training images are not necessary, which shows great flexibility in preparing training data. By applying RestainNet, all the images were mapped into the same staining style with CoNSeP dataset, as demonstrated in Fig.~\ref{fig:overview} (b).

\subsection{CNN Backbone}
To the best of our knowledge, HoVer-Net~\cite{graham2019hover} is the strongest baseline model for nuclear segmentation and classification. Thanks to the horizontal and vertical distance maps design, HoVer-Net is able to separate the overlapping nuclei. So in this challenge, HoVer-Net was employed as our baseline method, which was proposed to segment and classify nuclei in histology images simultaneously. The network consisted of three branches: Nuclear Pixel (NP), HoVer, Nuclear Classification (NC). NP was designed for nuclear semantic segmentation, and the HoVer learned the horizontal and vertical map of the nuclei. Combined with the results of NP, the markers were available for watershed algorithm that was used for post-processing. NC predicted the category of nuclear pixels, and merged the segmentation results of the above two branches to obtain the final results.
\begin{figure*}[t]
	\centering
	\includegraphics[width=.9\linewidth]{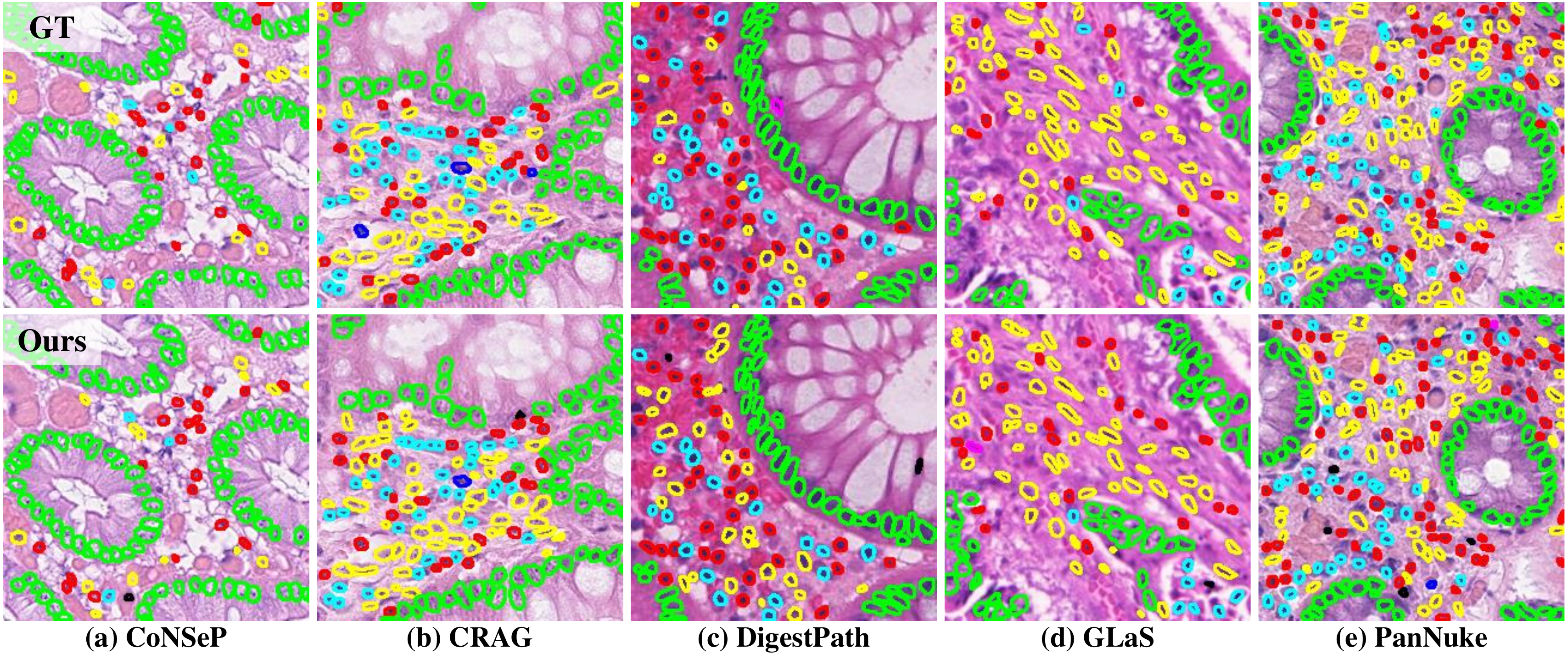}
	\caption{Qualitative results of different sub-datasets.}
	\label{fig:results}
\end{figure*}
\begin{table*}[t]
	\centering
	\caption{Quantitative Results - Development Set}
	\label{tab:development}
	\begin{tabular}{c|ccc|ccccccc}
		\toprule[2pt]
		\multicolumn{11}{c}{\textbf{Experiments on Different Configurations}}  \\ \hline
		\textbf{Model} &\textbf{Data} &\textbf{Stain} &\textbf{Loss} &\textbf{mPQ+} &\textbf{pla} &\textbf{neu} &\textbf{epi} &\textbf{lym} &\textbf{eos} &\textbf{con} \\
		\hline
		\multirow{9}{*}{HoVer-Net} & - & - & - &0.4735 &0.4960 &0.2121 &0.5936 &0.6704 &0.2928 &\textbf{0.5760} \\
		& $\checkmark$ & - & - &\textbf{0.4808} &0.4951 &\textbf{0.2458} &0.5921 &0.6664 &0.3160 &0.5694 \\
		& - & $\checkmark$ & - &0.4658 &0.4934 &0.1997 &0.5914 &0.6656 &0.2854 &0.5597 \\
		& - & - & $\checkmark$ &\textbf{0.4762} &0.4869 &0.2154 &\textbf{0.5946} &\textbf{0.6728} &0.3136 &0.5741 \\
		& $\checkmark$ & $\checkmark$ & - &0.4664 &0.4831 &0.2048 &0.5914 &0.6574 &0.3044 &0.5571 \\
		& - & $\checkmark$ & $\checkmark$ &0.4635 &0.4877 &0.1472 &0.5919 &0.6667 &\textbf{0.3271} &0.5605 \\
		& $\checkmark$ & - & $\checkmark$ &\textbf{0.4792} &\textbf{0.4969} &0.2318 &\textbf{0.5946} &0.6694 &0.3112 &0.5716 \\
		& $\checkmark$ & $\checkmark$ & $\checkmark$ &0.4642 &0.4794 &0.2024 &0.5904 &0.6591 &0.2997 &0.5543 \\
		\toprule[2pt]
		
		\multicolumn{11}{c}{\textbf{Cross-validation on Stain Normalization}}  \\ \hline
		\textbf{Model} &\textbf{Data} &\textbf{Stain} &\textbf{Loss} &\textbf{mPQ+} &\textbf{pla} &\textbf{neu} &\textbf{epi} &\textbf{lym} &\textbf{eos} &\textbf{con} \\
		\hline
		\multirow{5}{*}{HoVer-Net}		& - & - & - &\begin{tabular}[c]{@{}c@{}}0.4042\\ $\pm$ 0.0506\end{tabular} &\begin{tabular}[c]{@{}c@{}}0.4001\\ $\pm$ 0.0554\end{tabular} &\begin{tabular}[c]{@{}c@{}}0.1308\\ $\pm$ 0.1204\end{tabular} &\begin{tabular}[c]{@{}c@{}}\textbf{0.5368}\\ $\pm$ 0.0371\end{tabular} &\begin{tabular}[c]{@{}c@{}}0.5995\\ $\pm$ 0.0158\end{tabular} &\begin{tabular}[c]{@{}c@{}}\textbf{0.2676}\\ $\pm$ 0.1204\end{tabular} &\begin{tabular}[c]{@{}c@{}}0.4906\\ $\pm$ 0.0640\end{tabular} \\
		\cmidrule(lr){2-11} 
		
		&- & $\checkmark$ & - &\begin{tabular}[c]{@{}c@{}}0.4121\\ $\pm$ 0.0526\end{tabular} &\begin{tabular}[c]{@{}c@{}}0.4132\\ $\pm$ 0.0740\end{tabular} &\begin{tabular}[c]{@{}c@{}}\textbf{0.1762}\\ $\pm$ 0.1245\end{tabular} &\begin{tabular}[c]{@{}c@{}}0.5293\\ $\pm$ 0.0536\end{tabular} &\begin{tabular}[c]{@{}c@{}}\textbf{0.6006}\\ $\pm$ 0.0306\end{tabular} &\begin{tabular}[c]{@{}c@{}}0.2609\\ $\pm$ 0.1117\end{tabular} &\begin{tabular}[c]{@{}c@{}}0.4923\\ $\pm$ 0.0222\end{tabular} \\
		\cmidrule(lr){2-11}
		& $\checkmark$ & $\checkmark$ & $\checkmark$ &\begin{tabular}[c]{@{}c@{}}\textbf{0.4140}\\ $\pm$ 0.0424\end{tabular} &\begin{tabular}[c]{@{}c@{}}\textbf{0.4181}\\ $\pm$ 0.0759\end{tabular} &\begin{tabular}[c]{@{}c@{}}0.1688\\ $\pm$ 0.0943\end{tabular} &\begin{tabular}[c]{@{}c@{}}0.5340\\ $\pm$ 0.0520\end{tabular} &\begin{tabular}[c]{@{}c@{}}0.5987\\ $\pm$ 0.0323\end{tabular} &\begin{tabular}[c]{@{}c@{}}0.2619\\ $\pm$ 0.0647\end{tabular} &\begin{tabular}[c]{@{}c@{}}\textbf{0.5024}\\ $\pm$ 0.0222\end{tabular} \\
		\toprule[2pt]
	\end{tabular}
\end{table*}

\subsection{Loss Functions}
Besides the original losses introduced in HoVer-Net, we also introduced a cost-sensitive loss in the NC branch. Imbalanced data is a regular occurrence in target identification and classification tasks. The imbalance of different cell types is particularly pronounced in the CoNIC challenge. For example, epithelials, lymphocytes and connective cells contribute for a substantial portion of the ratio, whereas eosinophils and neutrophils account for a negligible portion. The imbalance of cell types will result in a decrease in segmentation and classification performance. We proposed a cost-sensitive loss function for multi-class nuclear segmentation and classification to address this problem. By incorporating the cost-sensitive matrix into the loss function, the impact of sample imbalance could be alleviated, and the classification accuracy of small categories can be improved.
For multi-class tasks, it is assumed that the number of categories is $\boldsymbol{N}$. The cost-sensitive matrix $\boldsymbol{M}$ with the size of $\boldsymbol{N}\times\boldsymbol{N}$ can be defined as follows:
\[
\boldsymbol{M} =
\begin{bmatrix}
m_{(0,0)} & m_{(0,1)} & \ldots & m_{(0, N-1)}\\
m_{(1,0)} & m_{(1,1)} & \ldots & m_{(1, N-1)}\\
\vdots & \vdots & \ddots & \vdots\\
m_{(N-1,0)} & m_{(N-1,1)} & \ldots & m_{(N-1, N-1)}\\
\end{bmatrix}
\]
where $ m_{(j,k)}$ represents the cost when class $k$ is incorrectly classified as class $j$. The diagonal element of matrix $\boldsymbol{M} $ is 0. The values of non-diagonal elements on the cost-sensitive matrix are set according to the ratio of the number of samples.

\section{Experimental Results}
We first demonstrate the results of both tasks on the leaderboard. Then we conduct a local development set to further evaluate the importance of each component in the proposed pipeline. Finally we demonstrate some qualitative results in Fig.~\ref{fig:results}. 

In the preliminary test phase, the quantitative results of both tasks are shown in Table~\ref{tab:task1}. Note that, we do not specifically design a regression model for the task 2, the results of the task 2 are directly generated from the results of the task 1. According to the results of the leaderboard, the mPQ+ and r2 of our proposed pipeline achieves 0.40665 (Rank 49th) and 0.62199 (Rank 10th), respectively. We can observe that both stain normalization and data augmentation can greatly benefit the nuclear segmentation and classification task for all the classes. Since each team can only submit the algorithm for once, we do not have chances to test all the configurations of different components. Therefore, we conduct a local development set of further evaluation.

Since task 2 is highly related to task 1. In this experiment, we only evaluate the proposed pipeline in nuclei segmentation and classification task. We split the Lizard dataset into a training set and a test set. All the models were trained for 100 epoches with no validation set. As demonstrated in the upper part of Table~\ref{tab:development}, HoVer-Net alone achieves moderate performance. Both Data augmentation and cost-sensitive loss improve the segmentation and classification performance. Specifically, the result of eosinophils, the one with the fewest labels, increases from 0.2928 to 0.3136 by cost-sensitive loss. Combining data augmentation and cost-sensitive loss can also get an obvious improvement for the minority classes, neutrophils and eosinophils. However, when equipped with stain normalization, the quantitative performance decreases in all the three models with it, which in contradiction with the leaderboard result.
The reason is that when we construct the local development set, we mix up the data from all five datasets. So the stain style of the test set if not blinded for the CNN model. In the meanwhile, GAN-based stain normalization model may cause information loss. So stain normalization in this experiment does not bring positive effect which is not as expected.

Next, we introduce a `5-fold cross validation-like' experiment to evaluate the effectiveness of stain normalization. In this experiment, each sub-dataset is regarded as one fold to ensure the testing fold is always blinded to the CNN model. The lower part of Table~\ref{tab:development} demonstrates the quantitative results of the segmentation and classification task. When the color styles are blinded to the model, stain normalization component generally help improve the results.

\section{Conclusion}
In this paper, we proposed a standardized nuclei segmentation and classification pipeline for CoNIC challenge. The pipeline integrates several pluggable components, including data augmentation, stain normalization, CNN backbone and loss functions. Experiments on the preliminary test set and the local development set demonstrate the effectiveness of this pipeline.

\section*{Acknowledgment}
This research was supported in part by National Science Fund for Distinguished Young Scholars, China (Grant No. 81925023), National Natural Science Foundation of China (Grant Nos. 62002082, 62102103),China Postdoctoral Science Foundation (Grant No. 2021M690753), 2020 National Innovation and Entrepreneurship Training Program for College Students (Grant No. 202010595023).


\bibliographystyle{IEEEtran}\bibliography{ref}

\begin{thebibliography}{10}
\providecommand{\url}[1]{#1}
\csname url@samestyle\endcsname
\providecommand{\newblock}{\relax}
\providecommand{\bibinfo}[2]{#2}
\providecommand{\BIBentrySTDinterwordspacing}{\spaceskip=0pt\relax}
\providecommand{\BIBentryALTinterwordstretchfactor}{4}
\providecommand{\BIBentryALTinterwordspacing}{\spaceskip=\fontdimen2\font plus
\BIBentryALTinterwordstretchfactor\fontdimen3\font minus
  \fontdimen4\font\relax}
\providecommand{\BIBforeignlanguage}[2]{{%
\expandafter\ifx\csname l@#1\endcsname\relax
\typeout{** WARNING: IEEEtran.bst: No hyphenation pattern has been}%
\typeout{** loaded for the language `#1'. Using the pattern for}%
\typeout{** the default language instead.}%
\else
\language=\csname l@#1\endcsname
\fi
#2}}
\providecommand{\BIBdecl}{\relax}
\BIBdecl

\bibitem{graham2021conic}
S.~Graham, M.~Jahanifar, Q.~D. Vu, G.~Hadjigeorghiou, T.~Leech, D.~Snead,
  S.~E.~A. Raza, F.~Minhas, and N.~Rajpoot, ``Conic: Colon nuclei
  identification and counting challenge 2022,'' \emph{arXiv preprint
  arXiv:2111.14485}, 2021.

\bibitem{bera2019artificial}
K.~Bera, K.~A. Schalper, D.~L. Rimm, V.~Velcheti, and A.~Madabhushi,
  ``Artificial intelligence in digital pathology—new tools for diagnosis and
  precision oncology,'' \emph{Nature reviews Clinical oncology}, vol.~16,
  no.~11, pp. 703--715, 2019.

\bibitem{zhao2020triple}
B.~Zhao, X.~Chen, Z.~Li, Z.~Yu, S.~Yao, L.~Yan, Y.~Wang, Z.~Liu, C.~Liang, and
  C.~Han, ``Triple u-net: Hematoxylin-aware nuclei segmentation with
  progressive dense feature aggregation,'' \emph{Medical Image Analysis},
  vol.~65, p. 101786, 2020.

\bibitem{graham2019hover}
S.~Graham, Q.~D. Vu, S.~E.~A. Raza, A.~Azam, Y.~W. Tsang, J.~T. Kwak, and
  N.~Rajpoot, ``Hover-net: Simultaneous segmentation and classification of
  nuclei in multi-tissue histology images,'' \emph{Medical Image Analysis},
  vol.~58, p. 101563, 2019.

\bibitem{qu2020weakly}
H.~Qu, P.~Wu, Q.~Huang, J.~Yi, Z.~Yan, K.~Li, G.~M. Riedlinger, S.~De,
  S.~Zhang, and D.~N. Metaxas, ``Weakly supervised deep nuclei segmentation
  using partial points annotation in histopathology images,'' \emph{IEEE
  transactions on medical imaging}, vol.~39, no.~11, pp. 3655--3666, 2020.

\bibitem{zhang2022ddtnet}
X.~Zhang, X.~Zhu, K.~Tang, Y.~Zhao, Z.~Lu, and Q.~Feng, ``Ddtnet: A dense
  dual-task network for tumor-infiltrating lymphocyte detection and
  segmentation in histopathological images of breast cancer,'' \emph{Medical
  Image Analysis}, p. 102415, 2022.

\bibitem{graham2021lizard}
S.~Graham, M.~Jahanifar, A.~Azam, M.~Nimir, Y.-W. Tsang, K.~Dodd, E.~Hero,
  H.~Sahota, A.~Tank, K.~Benes \emph{et~al.}, ``Lizard: A large-scale dataset
  for colonic nuclear instance segmentation and classification,'' in
  \emph{Proceedings of the IEEE/CVF International Conference on Computer
  Vision}, 2021, pp. 684--693.

\bibitem{gamper2019pannuke}
J.~Gamper, N.~Alemi~Koohbanani, K.~Benet, A.~Khuram, and N.~Rajpoot, ``Pannuke:
  an open pan-cancer histology dataset for nuclei instance segmentation and
  classification,'' in \emph{European Congress on Digital Pathology}.\hskip 1em
  plus 0.5em minus 0.4em\relax Springer, 2019, pp. 11--19.

\bibitem{cheng2021deep}
J.~Cheng, Z.~Wang, Z.~Liu, Z.~Feng, H.~Wang, and X.~Pan, ``Deep adversarial
  image synthesis for nuclei segmentation of histopathology image,'' in
  \emph{2021 2nd Asia Conference on Computers and Communications (ACCC)}.\hskip
  1em plus 0.5em minus 0.4em\relax IEEE, 2021, pp. 63--68.

\bibitem{zhao2022restainnet}
B.~Zhao, J.~Lin, C.~Liang, Z.~Yi, X.~Chen, B.~Li, W.~Qiu, D.~Li, L.~Liang,
  C.~Han \emph{et~al.}, ``Restainnet: a self-supervised digital re-stainer for
  stain normalization,'' \emph{arXiv preprint arXiv:2202.13804}, 2022.

\bibitem{zhou2005training}
Z.-H. Zhou and X.-Y. Liu, ``Training cost-sensitive neural networks with
  methods addressing the class imbalance problem,'' \emph{IEEE Transactions on
  knowledge and data engineering}, vol.~18, no.~1, pp. 63--77, 2005.

\end{thebibliography}

\end{document}